\documentclass[12pt]{iopart}
\pdfoutput=1 
\usepackage[utf8]{inputenc}
\usepackage{graphicx}
\usepackage[position=top,skip=1pt,singlelinecheck=false]{subfig}	
\usepackage{textgreek}
\def \ket#1{{|\,#1\,\rangle}}
\def \bra#1{{\langle\,#1\,|}}
\def \cu#1{{\hat{c}_{#1\uparrow}}}
\def \cd#1{{\hat{c}_{#1\downarrow}}}

\def \cud#1{{\hat{c}_{#1\uparrow}^\dagger}}
\def \cdd#1{{\hat{c}_{#1\downarrow}^\dagger}}

\def \nu#1{{\hat{n}_{#1\uparrow}}}
\def \nd#1{{\hat{n}_{#1\downarrow}}}
 
\begin{document}

\title{Dynamics of Ultracold Quantum Gases in the Dissipative Fermi-Hubbard Model}

\author{K~Sponselee$^1$, L~Freystatzky$^{1,3}$, B~Abeln$^1$, M~Diem$^1$, B~Hundt$^1$, A~Kochanke$^1$, T~Ponath$^1$, B~Santra$^1$, L~Mathey$^{1,2,3}$, K~Sengstock$^{1,2,3}$ and C~Becker$^{1,2}$}
\address{$^1$ Zentrum f\"ur Optische Quantentechnologien, Universit\"at Hamburg, Luruper Chaussee 149, 22761 Hamburg, Germany}
\address{$^2$ Institut f\"ur Laserphysik, Universit\"at Hamburg, Luruper Chaussee 149, 22761 Hamburg, Germany}
\address{$^3$ The Hamburg Centre for Ultrafast Imaging, Luruper Chaussee 149, 22761 Hamburg, Germany}
\ead{cbecker@physnet.uni-hamburg.de}

\maketitle

\begin{abstract}
We employ metastable ultracold $^{173}$Yb atoms to study dynamics in the 1D dissipative Fermi-Hubbard model experimentally and theoretically,
and observe a complete inhibition of two-body losses after initial fast transient dynamics. 
We attribute the suppression of particle loss to the dynamical generation of a highly entangled Dicke state. 
For several lattice depths and for two- and six-spin component mixtures we find very similar dynamics, showing that the creation of strongly correlated states is a robust and universal phenomenon.
This offers interesting opportunities for precision measurements.
\end{abstract}

\section{Introduction}
\label{sec:introduction}

Real world quantum systems are inherently coupled to an environment, causing decoherence and dissipation. 
This is often regarded as one of the biggest impediments to the investigation and utilization of quantum systems.
Counterintuitively, it was recently proposed to employ engineered dissipation for quantum state preparation \cite{Diehl2008, Verstraete2009, Diehl2011,Foss-Feig2012}, entropy transfer \cite{Griessner2006}, or as a tool for the measurement of correlations \cite{Baur2010}, to name only a few examples.

Cold atom experiments are commonly appreciated for the high degree of control over most system parameters and their strong decoupling from the environment, which makes them very well-suited as quantum simulators of interacting quantum many-body systems \cite{Gross2017}. 
Furthermore, ultracold quantum gases are ideal candidates for studying the influence of dissipation under well-controlled conditions by carefully introducing a certain coupling to a bath \cite{Catani2009,Chen2014} or tailored particle losses. 
One-body losses were implemented in different ways, as for example with an electron beam pointing at a Bose-Einstein condensate of $^{87}$Rb atoms, where a suppression of losses due to the continuous quantum Zeno effect \cite{Barontini2013} and a dissipatively driven non-equilibrium phase transition was observed  \cite{Labouvie2015, Labouvie2016}. 
In other experiments, near resonant photon scattering was used to study the thermalization of a localized many body state in the presence of dissipation \cite{Luschen2017}, or measurement induced suppression of tunneling \cite{Patil2015}. 
Two-body losses were realized in the form of inelastic collisions. 
For a system of bosonic $^{87}$Rb Feshbach-molecules in an array of one-dimensional tubes a suppression of particle losses was studied \cite{Syassen2008} and theoretically analyzed \cite{Garcia-Ripoll2009,Durr2009}. 
Loss suppression was observed in a similar system with hetero-nuclear fermionic molecules \cite{Yan2013,Zhu2014}. 
Tuning the inelastic two-body loss rate via a photo-association resonance in a system of bosonic $^{174}$Yb atoms, a delay of the  melting of the Mott insulating state was found \cite{Tomita2017}. 
In a gas of Cs atoms the formation of a metastable Mott insulator with attractive interactions was observed when increasing the three-body loss rate employing a magnetic Feshbach resonance \cite{Mark2012}.

Here we report on the realization of the one-dimensional dissipative Fermi-Hubbard model with two- and six-spin component mixtures of ultracold $^{173}$Yb atoms in the metastable-state $^{3}$P$_{0}$ in an array of one-dimensional lattice systems.
Dissipation naturally occurs in the form of inelastic two-body collisions. 
We work in a regime where the inelastic interaction $ \hbar \Gamma_{0}$ is comparable to the on-site interaction $U$, which has not been studied in a fermionic system yet.

We find that after initial loss dynamics on the time scale of superexchange particle loss ceases and a considerable fraction of atoms remains. 
For two- and six-spin component mixtures we observe universal dynamics in the sense that the remaining fraction does not depend on the lattice depth. 
This can be explained by the emergence of a highly correlated state which is predicted \cite{Foss-Feig2012} to uniquely be a Dicke state \cite{Radiation1954}.
Dicke states are characterized by a fully symmetric spin wave function with maximal total angular momentum $S_\mathrm{eff} = S_\mathrm{max}$ and a minimal uncertainty $\Delta S_z$ (as depicted in \fref{fig:bloch_sphere}), and notably exhibit optimal metrological properties \cite{Lucke2014,Apellaniz2015}.

\section{Model}
\label{sec:model}


\begin{figure}[b]
  \subfloat[\label{fig:bloch_sphere}]
  {
    \includegraphics[width=0.40\textwidth]{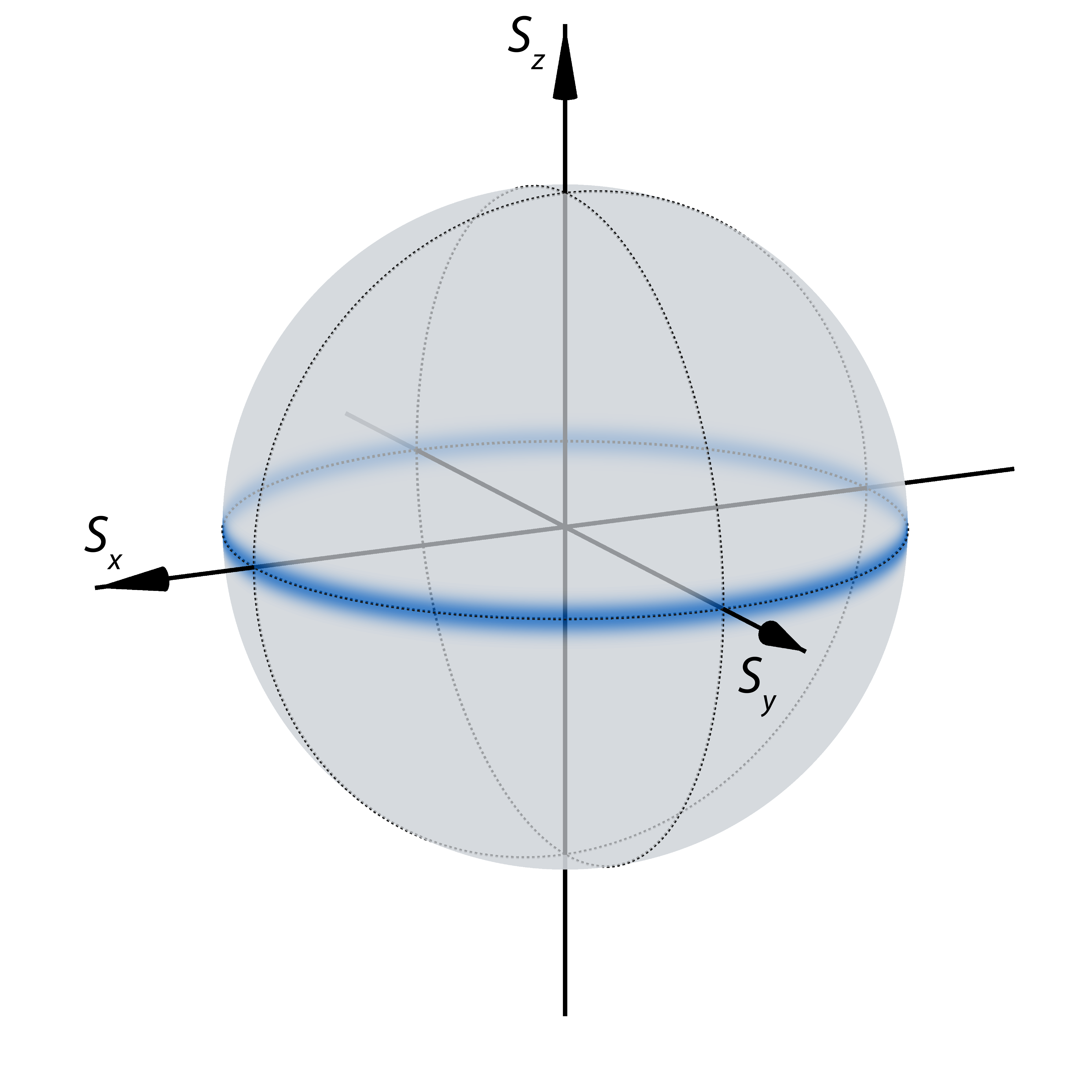}
  }
  \hfill
  \subfloat[\label{fig:g2_configurations}]
  {
    \includegraphics[width=0.50\textwidth]{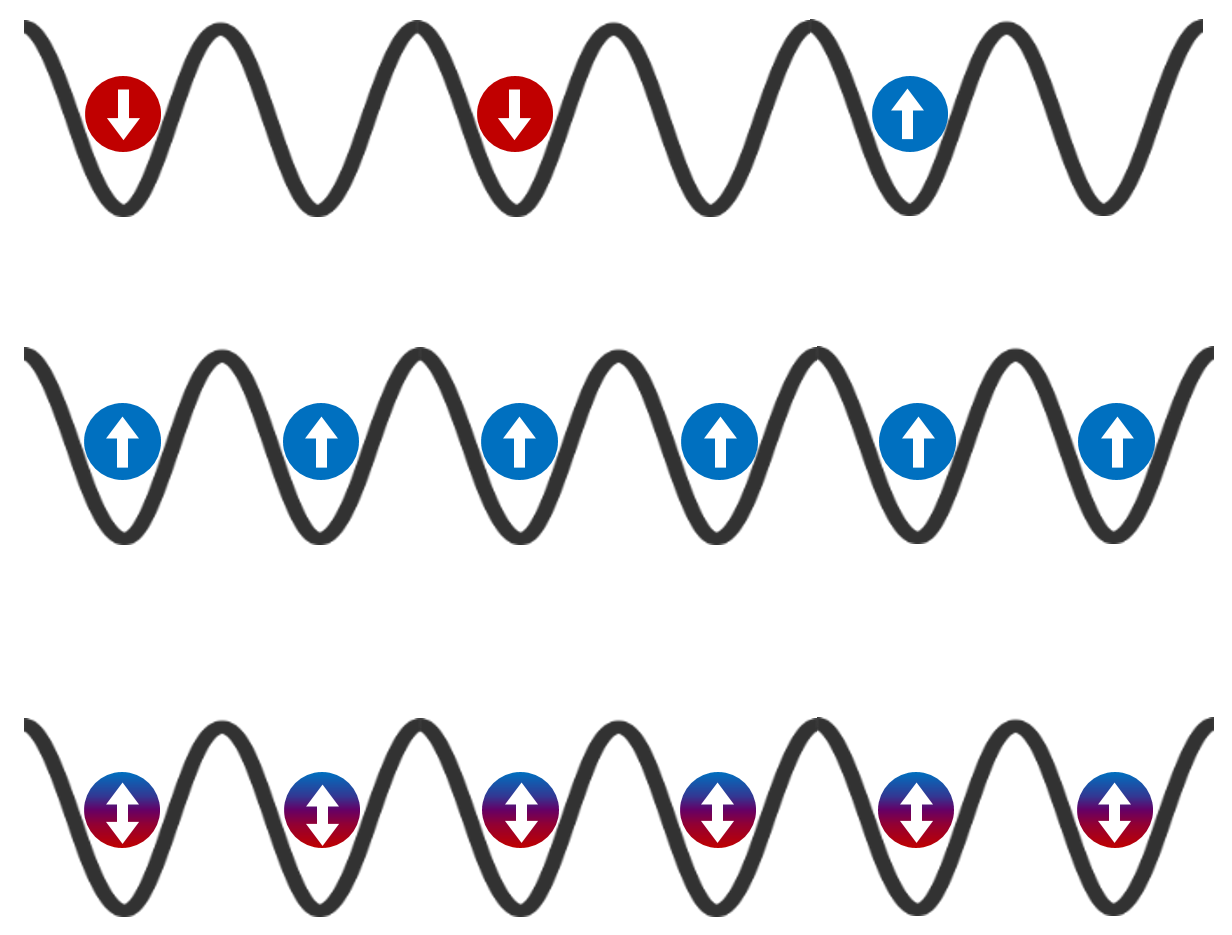}
  }
  \caption{(a) Sketch of a Dicke state as a thin blue ring around the equator on a generalized Bloch sphere, whose radius is given by $S~=~S_\mathrm{max}$. The uncertainty $\Delta S_z$ is minimal. 
     (b) Different configurations exhibiting $g^{(2)} = 0$. (top) a state without nearest neighbours, (centre) a ferromagnetic state, (bottom) a sketch of a highly entangled Dicke state.}
     \label{fig:dicke_states_scetches}
\end{figure}

To gain insight into the expected dynamical behaviour of our system we consider the dissipative Fermi-Hubbard model in one dimension. 
It is characterized by three quantities, the nearest neighbour tunneling matrix element $J$, the on-site interaction energy $U~=~4\pi\hbar^2a_{ee}m^{-1}\int|w_0(\bi{r})|^4\rmd\bi{r}$ and the on-site loss rate $\Gamma_0~=~\hbar\beta_{ee}~\int|w_0(\bi{r})|^4\rmd\bi{r}$ \cite{Yan2013}. 
Here $\hbar$ is the reduced Planck constant, $a_{ee}~\approx~306.2~a_0$ is the elastic $s$-wave scattering length in units of the Bohr radius $a_0$ \cite{Scazza2014,Kitagawa2008}, $m$ is the mass, $w_0(\bi{r})$ is the Wannier function, and $\beta_{ee}~\approx~2.2\cdot10^{-11}$~cm$^3$~s$^{-1}$ is the two-body loss coefficient \cite{Scazza2014}.

\begin{figure}[t]
    \centering
    \begin{minipage}[t]{0.30\textwidth}
    \subfloat[\label{fig:system_schematic_a}]
    {
      \includegraphics[width=\textwidth]{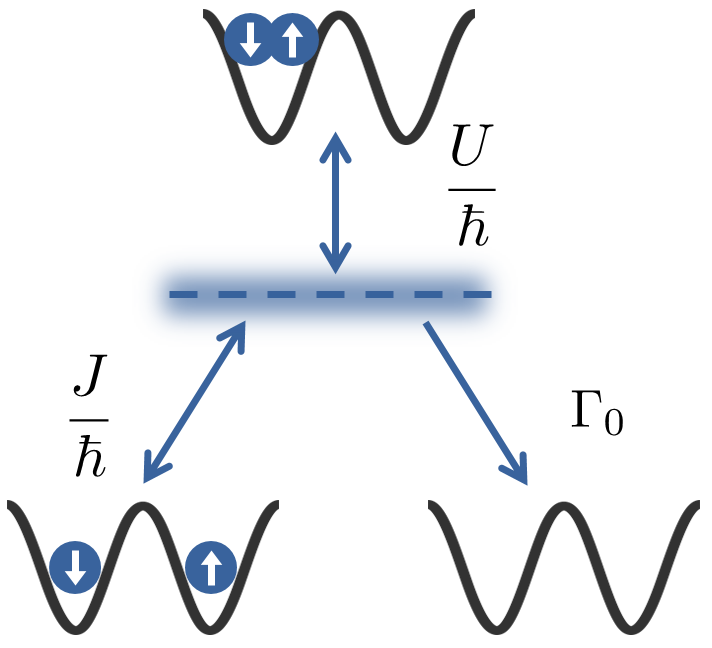}
    }
    
    \subfloat[\label{fig:system_schematic_b}]
    {
      \includegraphics[width=\textwidth]{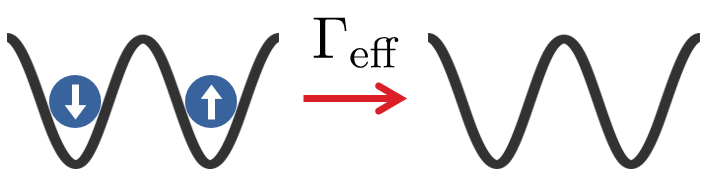}
    }    
    \end{minipage}
    \hfill
    \begin{minipage}[t]{0.55\textwidth}
    \subfloat[\label{fig:gamma_eff_plot}]
    {
      \includegraphics[width=\textwidth]{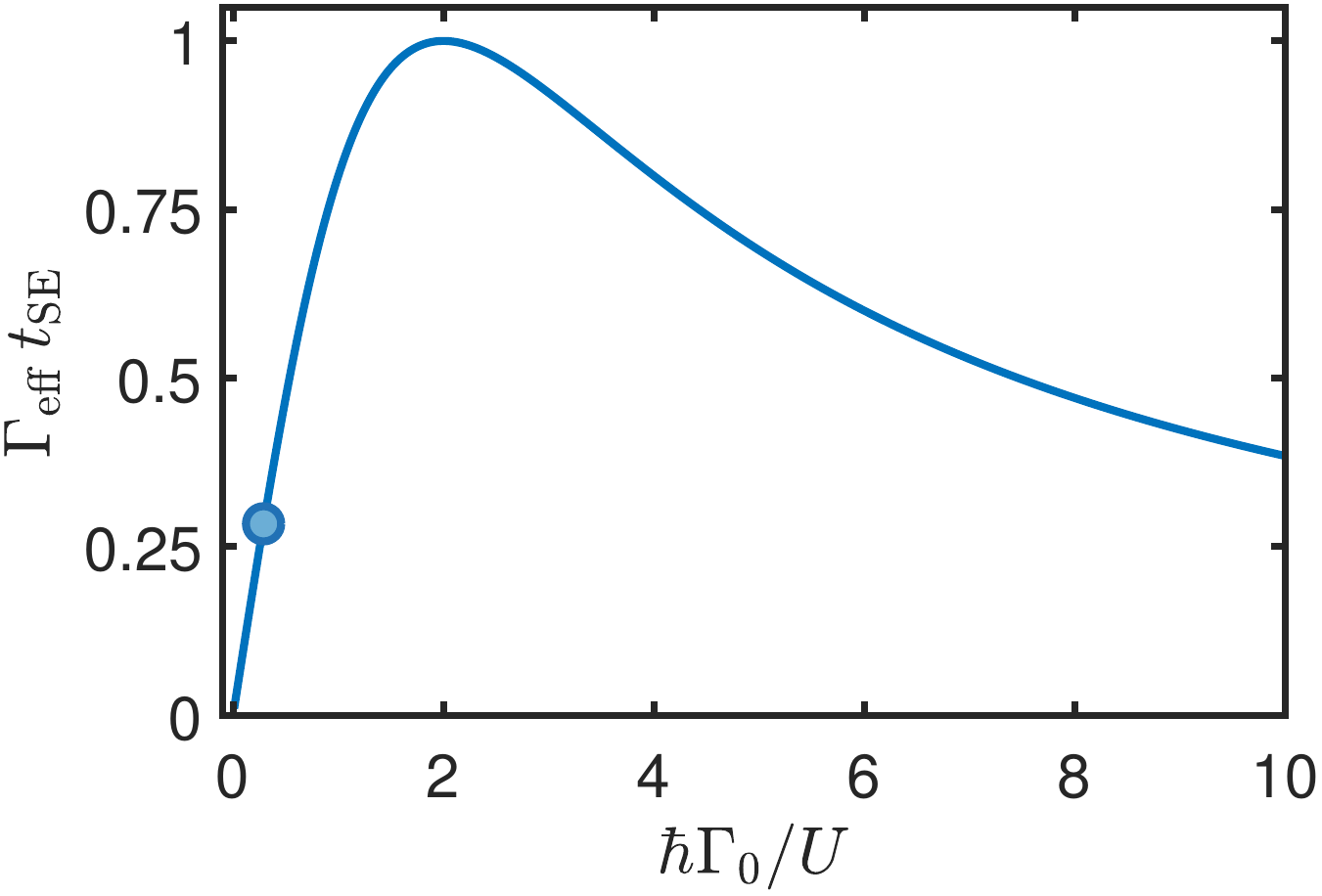}
    }
    \end{minipage}
    \caption{(a) Schematic of a simplified two well model reminiscent of the dissipative Fermi-Hubbard model with the tunneling $J$, on-site interaction energy $U$ and on-site loss rate $\Gamma_0$. (b) Schematic of the effective two-level system. (c) Effective loss rate as a function of dissipation strength $\hbar \Gamma_0/U$. The point indicates the fixed dissipation strength $\hbar \Gamma_0 / U = 0.29$ for $^{173}$Yb.}
    \label{fig:system_description_all}
\end{figure}

For the full description of the dynamics we have performed numerical simulations of the underlying master equation (see \ref{sec_app:master_equation}).
Regarding our experimentally accessible quantity of interest, the particle number dynamics $N(t)$, the essential physics can already be understood by considering a simplified model \cite{Garcia-Ripoll2009,Yan2013,Zhu2014}. 
In this simplified model, depicted in \fref{fig:system_schematic_a}, the intermediate doubly occupied state can be integrated out provided the tunneling is the smallest energy scale of the system ($J \ll U, \hbar \Gamma_0$).
The dynamics in the resulting reduced two-level system (shown in \fref{fig:system_schematic_b}) is governed by an effective loss rate \cite{Garcia-Ripoll2009,Yan2013,Zhu2014}
\begin{equation}
\label{eq:gamma_eff}
 \Gamma_\textrm{eff} = 4 \, \frac{J^2}{\hbar U} \left( \frac{U/\hbar \Gamma_0}{1+4(U/\hbar \Gamma_0)^2} \right),
\end{equation}
shown in \fref{fig:gamma_eff_plot}. 
We identify two regions where losses are suppressed. 
For $U \gg \hbar \Gamma_0$, in the Mott insulator regime, losses are strongly reduced caused by a suppression of double occupancy by elastic interactions. 
For $\hbar \Gamma_0 \gg U$, in the quantum Zeno regime, losses are suppressed because the strong inelastic two-body interactions act as a continuous measurement of doubly occupied sites, prohibiting the time evolution of the system. 

%

$U$ and $\Gamma_0$ both scale with the Wannier function, and thus for $^{173}$Yb the dimensionless dissipation strength $\hbar \Gamma_0 / U = 0.29$ cannot be tuned by changing the lattice depth and is thus constant in the measurements presented here.
Our system is therefore in the largely unexplored regime where $U$ and $\hbar \Gamma_0$ are comparable.

Using this effective loss rate, the time evolution of the total particle number can be described by \cite{Garcia-Ripoll2009}
\begin{equation}
\label{eq:diffential_N}
 \frac{\rmd N(t)}{\rmd t} = - \mathcal{P}_\mathcal{N} \frac{\kappa}{N_0} N(t)^2,
\end{equation}
where $\mathcal{P}_\mathcal{N} = (\mathcal{N}-1)/\mathcal{N}$ is the probability of finding an atom in a different spin state on a neighbouring lattice site for an uncorrelated spin mixture with $\mathcal{N}$ spin components at unitary filling (see \ref{sec_app:general_2body_loss_eq}). 
$N_0 \equiv N(t=0)$ is the initial number of atoms and $\kappa$ is the loss coefficient, which is related to the effective loss rate through \cite{Yan2013}
\begin{equation}
\label{eq:kappa}
 \kappa = 4 \, q \, \Gamma_\textrm{eff} \, g^{(2)} \, \eta_0,
\end{equation}
where $q$ is the number of nearest neighbours ($q=2$ for a 1D lattice), $\eta_0$ is the initial filling of the lattice ($\eta_0~=~1$ for a lattice with unitary filling), and $g^{(2)}$ is the nearest neighbour correlation function. 
For two-spin components $g^{(2)}$ has the form \cite{Baur2010}
\begin{equation}
\label{eq:g2}
 g^{(2)} = \frac{1}{N-1} \sum_{\left<ij\right>} \frac{\left< \hat{n}_i \hat{n}_j - 4 \hat{\bi{S}}_i \cdot \hat{\bi{S}}_j  /\hbar^2 \right>}{\left<\hat{n}_i\right>\left<\hat{n}_j\right>},
\end{equation}
where $\hat{n}_i$ and $\hat{\bi{S}}_i$ are the number and spin operators for lattice site $i$, respectively. 
Here $\left<ij\right>$ indicates the summation over nearest neighbours. 

Ultracold quantum gases offer the intriguing possibility to study high-spin systems \cite{Taie2010,Krauser2012,Stamper-Kurn2013,Krauser2014,Pagano2014,Hofrichter2016}, i.e. systems which are composed of particles with a large intrinsic spin $S>1/2$.
In our case, $^{173}$Yb possesses $S=5/2$, which together with the underlying SU$(6)$ symmetry allows working with up to six different spin components. 
Therefore we generalize the required nearest neighbour correlation function for an arbitrary number of spin components and find
\begin{equation}
g^{(2)}= \frac{1}{N-1} \sum_{\left<ij\right>} \sum\limits_{\sigma\neq\sigma'} \frac{ \left< \hat{n}_{i\sigma} \hat{n}_{j\sigma'}  +  \hat{n}_{i\sigma'} \hat{n}_{j\sigma}  -  \hat{c}_{i\sigma'}^\dagger \hat{c}_{j\sigma}^\dagger \hat{c}_{i\sigma}^{} \hat{c}_{j\sigma'}^{} - \hat{c}_{i\sigma}^\dagger \hat{c}_{j\sigma'}^\dagger \hat{c}_{i\sigma'}^{} \hat{c}_{j\sigma}^{} \right> }{ \left< \hat{n}_i \right> \left< \hat{n}_j \right>},
\end{equation}
where $\hat{n}_{i\sigma} \equiv \hat{c}_{i\sigma}^\dagger \hat{c}_{i\sigma}^{}$ is the number operator for lattice site $i$ and spin $\sigma$, and $\hat{c}_{i\sigma}^\dagger$ ($\hat{c}_{i\sigma}$) is the fermionic creation (annihilation) operator.

In general, the nearest neighbour correlation function as used here, by construction yields $g^{(2)}=1$ for a state without spin correlations.
A fully antisymmetric spin state, e.g. a quantum antiferromagnetic state, has $g^{(2)}=2$, while $g^{(2)}=0$ for a fully symmetric spin state. 
Several configurations result in a vanishing $g^{(2)}$ (depicted in \fref{fig:g2_configurations}), such as a state without nearest neighbours, a ferromagnetic state, and a highly entangled Dicke state (see \ref{sec_app:g2_Dicke_State}).
Equation \eref{eq:kappa} clearly shows that the build-up of any nearest neighbour correlations is directly reflected in the loss dynamics.

\section{Methods}
\label{sec:methods}

To realize the 1D dissipative Fermi-Hubbard model experimentally, we use the alkaline-earth-like fermionic isotope $^{173}$Yb in the metastable triplet ground-state $^3$P$_0$. 
$s$-wave collisions in the $^3$P$_0$ metastable-state are significantly inelastic ($\hbar \Gamma_0/U = 0.29$), leading to the desired two-body loss mechanism. 
Furthermore, the complete decoupling of the purely nuclear spin $F=5/2$ from the electronic degrees of freedom leads to SU$(\mathcal{N})$ symmetry of low energy interactions with $\mathcal{N}$ up to six.

In our experiment we use an efficient two-step laser cooling scheme based on a 2D and 3D magneto-optical trap before loading the atoms into a crossed optical dipole trap \cite{Dorscher2013}. 
We apply optical pumping on the intercombination transition $^1$S$_0 \rightarrow ^3$P$_1$ to achieve spin-polarized samples or two-spin mixtures. 
Subsequently, we evaporatively cool the atoms to quantum degeneracy. 
In this way, we create samples of $N_0 \approx 10^4$ atoms at $T \approx 0.25~T_\mathrm{F}$ where $T$ is the temperature and $T_\mathrm{F}$ is the Fermi temperature. 
Next, we ramp up a deep triangular 2D optical lattice ($V_\mathrm{lat}\approx42~E_\mathrm{r}$, where $E_\mathrm{r}$ is the recoil energy) and 1D lattice ($V_\mathrm{lat}\approx50~E_\mathrm{r}$) in the third direction to prepare a Mott insulator of $^1$S$_0$ ground-state atoms.
We then transfer the atoms to the metastable-state via a rapid-adiabatic passage on the ultra-narrow optical clock transition $^1$S$_0 \rightarrow ^3$P$_0$. 
We initiate the dissipative dynamics by rapidly ramping down the 1D lattice in $300$ \textmu s to variable final lattice depths of 5, 6 or $8~E_\mathrm{r}$, enabling tunneling in one dimension. 
In this way we create an array of effective 1D Fermi-Hubbard systems ($\hbar \omega_\perp \gg E_\mathrm{F}$, where $\omega_\perp$ is the radial harmonic confinement and $E_\mathrm{F}$ is the Fermi energy) in which we count both the number of ground-state atoms $^1$S$_0$ and metastable-state atoms $^3$P$_0$ after a variable holding time. 
Throughout the experiment, a homogeneous magnetic field with $B_{0}\approx3$~G is applied perpendicular to the long axis of the 1D lattice to provide a quantization axis for the atoms (referred to as the $z$-direction). 

We normalize our spin-mixture measurements to equalize the effect of the finite lifetime due to one-body losses, which are caused by collisions with background atoms, light scattering from the optical lattice, as well as radiative decay into the ground-state $^3$P$_0 \rightarrow ^1$S$_0$.
Since we are only interested in loss dynamics caused by two-body processes, we determine the one-body loss rate from measurements on non-interacting spin polarized samples.
Spin polarized fermions are not subject to two-body losses due to the Pauli exclusion principle. 
We quantify the one-body loss by fitting a decaying exponential to the spin-polarized measurements. 
In this way we find that the one-body loss timescales are about two orders of magnitude slower than the corresponding dissipative two-body dynamics in the 1D system for a given lattice depth, justifying this procedure.

To complement our measurements we numerically integrate a six-site model for two-spin components, using an Adam's method.
We initialize the system in an incoherent superposition of pure states without double occupancy, such that the density matrix is diagonal.
The initial weights of the pure states can be adjusted to tailor initial states with different correlations, while still ensuring normalization of the initial density matrix (see \ref{sec_app:master_equation}).
We study the atom number dynamics and obtain $g^{(2)}(t)$ from the full state information.

\section{Results}
\label{sec:results}

\begin{figure}[b]
  \centering
    \includegraphics[width=0.70\textwidth]{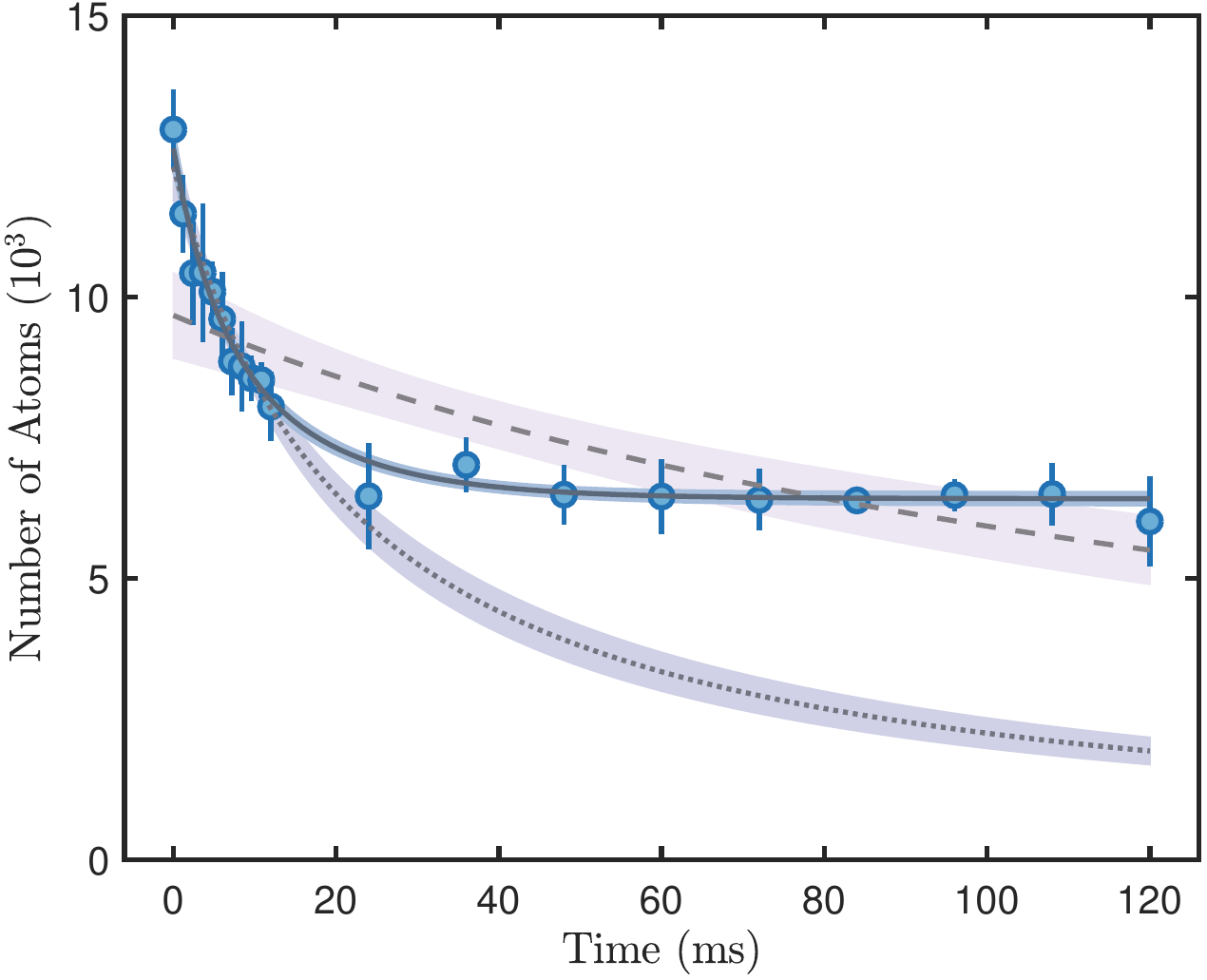}
  \caption{Measurement of the atom number as a function of time in a $6~E_\mathrm{r}$ deep 1D lattice. The experimental data is five times averaged and the error bars represent one standard deviation. The curves show different fits and the shaded areas show the 95\% confidence interval of the fit. The dashed line shows the fit with constant $\kappa$ to all data, the dotted line shows the fit to the first 11 data points with constant $\kappa$, and the solid line shows the fit of the form $\kappa(t) = \kappa_0 \, \rme^{-t/\tau}$ (see text). Here $\hbar/J~\approx~10$~ms and $t_\mathrm{SE}~\approx~38$~ms. 
  \label{fig:fitted_data_2S_6ER}}
\end{figure}

\Fref{fig:fitted_data_2S_6ER} shows a typical normalized measurement series of the number of atoms $N(t)$ as a function of time for a two-spin mixture with a final 1D lattice depth of $6~E_\mathrm{r}$. 
We observe a quick initial loss for times shorter than approximately 40~ms. 
For larger times, we find that the number of atoms stays constant within experimental errors. 

In order to properly fit the model \eref{eq:diffential_N} to our measurements, we have to assume an exponentially decaying $\kappa(t) = \kappa_0\,\rme^{-t/\tau}$. Here $\kappa_0 = 4 q \Gamma_\textrm{eff} g_0^{(2)} \eta_0 $ is the initial loss coefficient with the initial nearest neighbour correlation function $g_0^{(2)}$, and $\tau$ describes the timescale on which correlations build up. 
The solution to \eref{eq:diffential_N} yields
\begin{equation}
\label{eq:solution_exp_dec_kappa}
 N(t) = \frac{N_0}{1 + \mathcal{P}_\mathcal{N} \, \kappa_0 \, \tau - \mathcal{P}_\mathcal{N} \, \kappa_0 \, \tau \, \rme^{-t/\tau} } \stackrel{t\rightarrow \infty}{=} \frac{N_0}{1 + \mathcal{P}_\mathcal{N} \, \kappa_0 \, \tau},
\end{equation}
and importantly settles at a finite particle number for $t\rightarrow \infty$.
As seen in \fref{fig:fitted_data_2S_6ER}, we obtain excellent agreement with our experimental results, indicating that $g^{(2)}$ decays to zero on experimentally relevant timescales.

\begin{figure}[b]
    \subfloat[\label{fig:Theory_atom_number_and_g2}]
    {
      \includegraphics[width=0.45\textwidth]{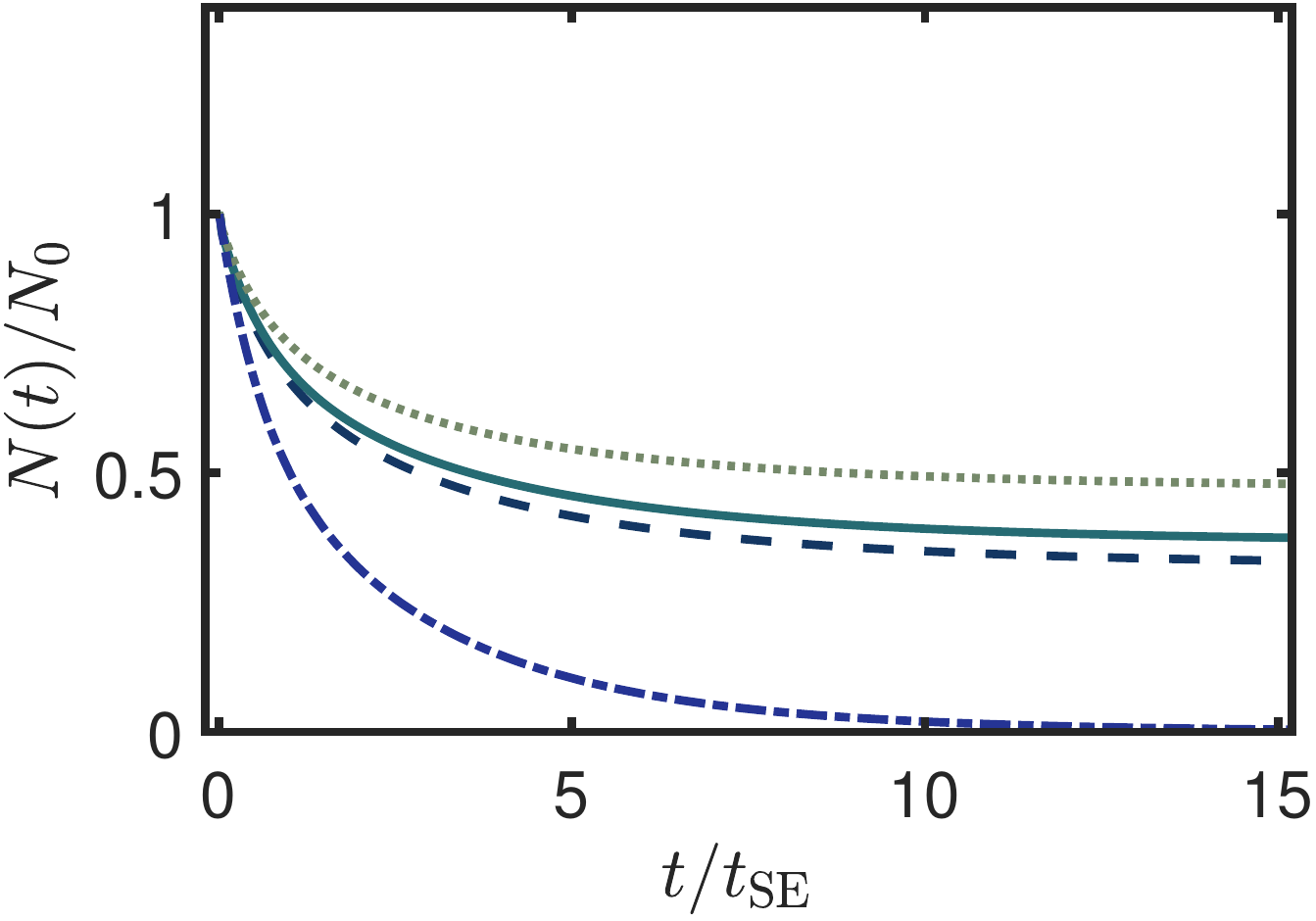}\llap{\makebox[3.3cm][l]{\raisebox{2.6cm}{\includegraphics[width=0.189\textwidth]{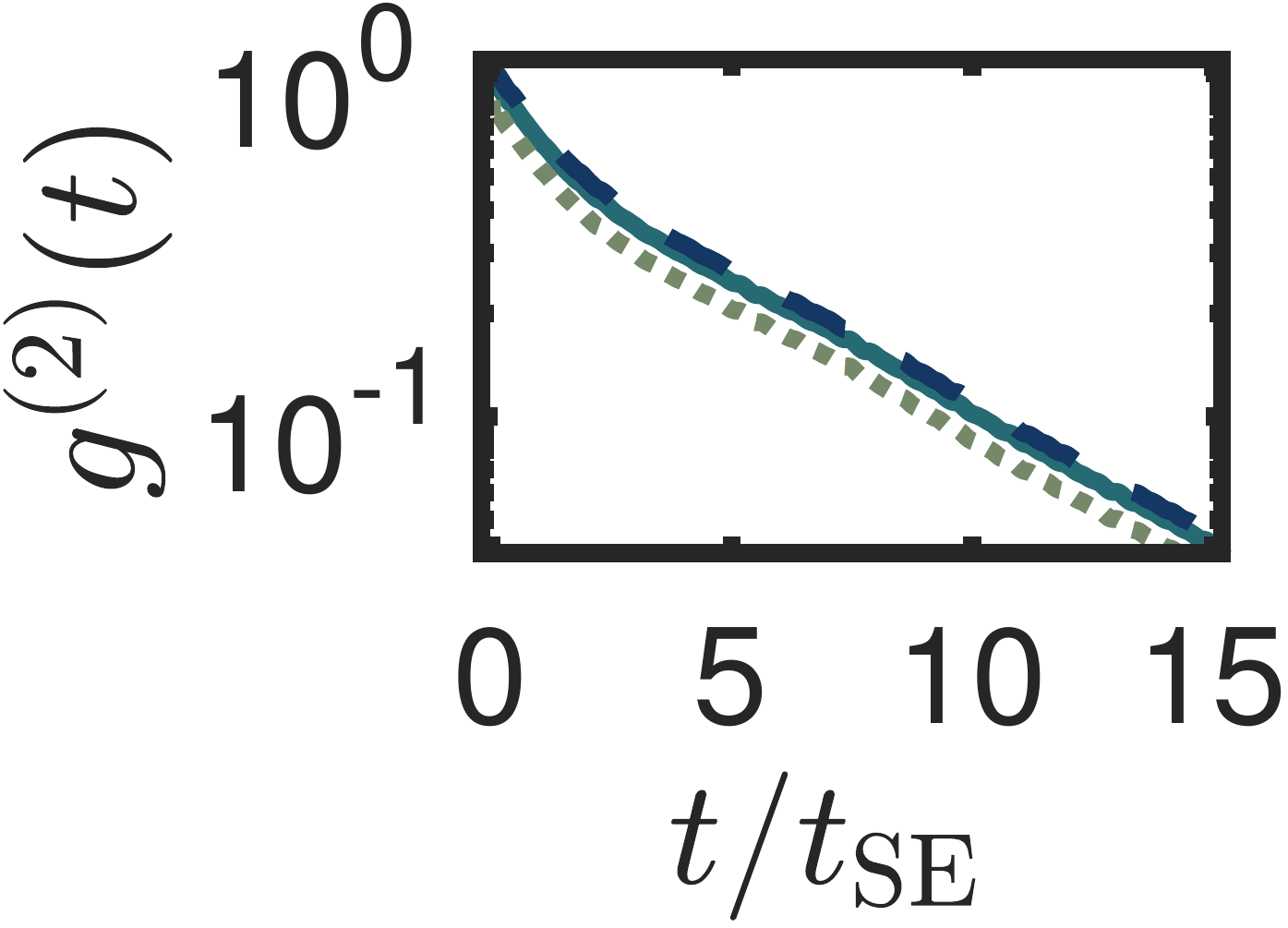}}}}
    }
    \hfill
    \subfloat[\label{fig:Theory_Comp_vary_U}]
    {
      \includegraphics[width=0.45\textwidth]{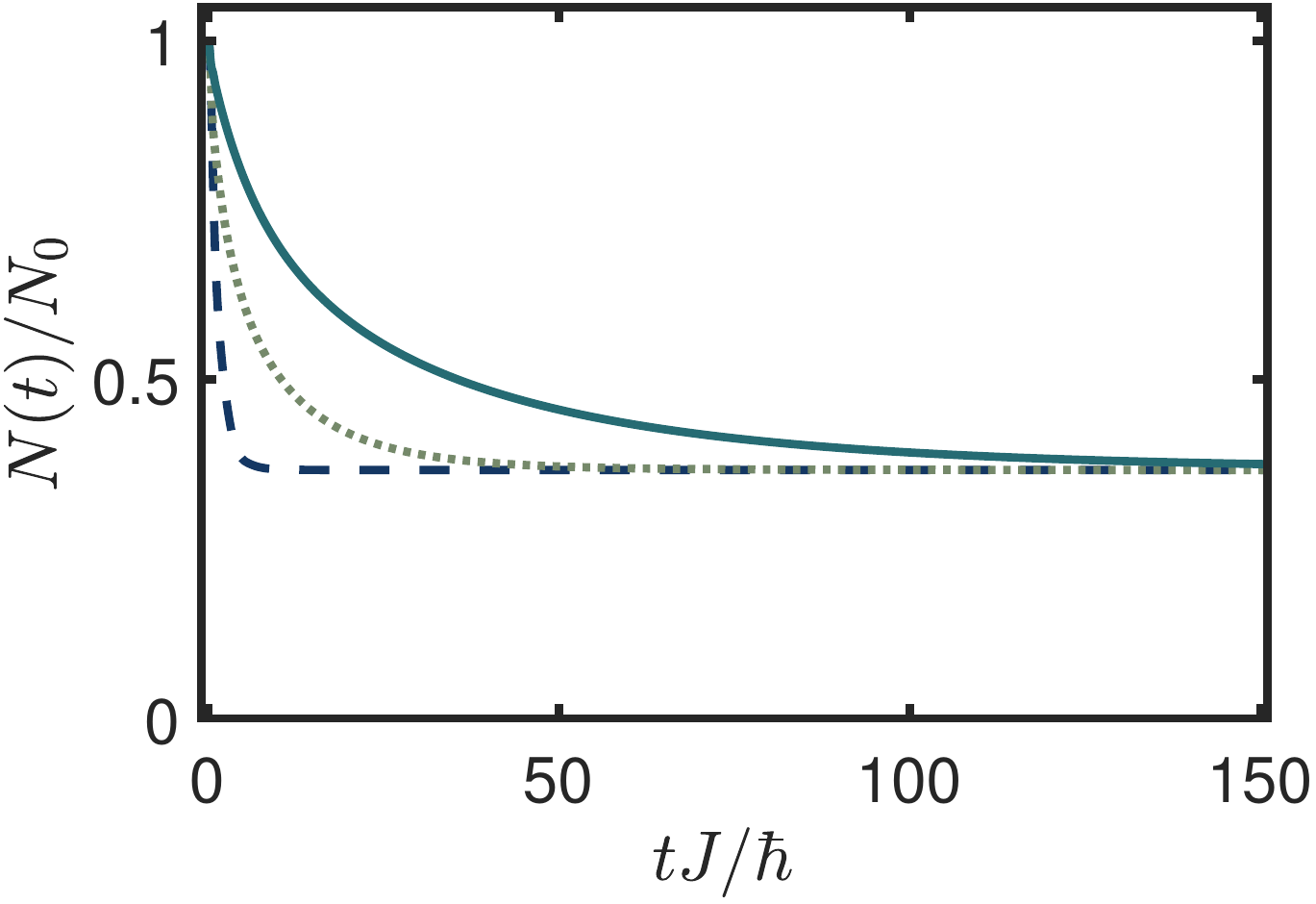}
    }
    \caption{Six-site simulation of a two-spin component Fermi gas with $\hbar \Gamma_0 = 3 J$. (a) Normalized atom number $N(t)/N_0$ as a function of dimensionless time $t/t_\mathrm{SE}$ for $\hbar \Gamma_0/U = 0.3$. The solid line shows results for an uncorrelated initial state, the dashed line shows results for smaller initial Dicke state weight, the dotted line shows results for larger initial Dicke state weight and the dash-dotted line shows results for a quantum antiferromagnetic state (see \ref{sec_app:master_equation}). The inset shows $g^{(2)}$ as a function of time. (b) Normalized atom number $N(t)/N_0$ as a function of $t J/\hbar$ for different on-site interactions $U$. In addition to $U = 0$ (dashed line), we studied $\hbar \Gamma_0/U = 0.6$ (dotted line) and $\hbar \Gamma_0/U = 0.3$ (solid line).}
    \label{fig:theory_results}
\end{figure}

We note that only an exponentially decaying $\kappa$ describes our data properly. 
As shown in \fref{fig:fitted_data_2S_6ER} fits with constant $\kappa$, either to the whole data set or exclusively to the initial dynamics, as was done in \cite{Syassen2008}, do not describe our data correctly.
Moreover, for a power-law dependent $\rmd \kappa / \rmd t \propto -\kappa^n$ we find that the final atom number always approaches zero for any power $n>1$.
Our simulations of the full master equation for different initial states also reveal an approximately exponentially decreasing $g^{(2)}$, strongly supporting our assumptions (see \fref{fig:Theory_atom_number_and_g2}). 


\begin{figure}[b]
  \subfloat[\label{fig:norm_losses_568Er_2S}]
  {
    \includegraphics[width=0.45\textwidth]{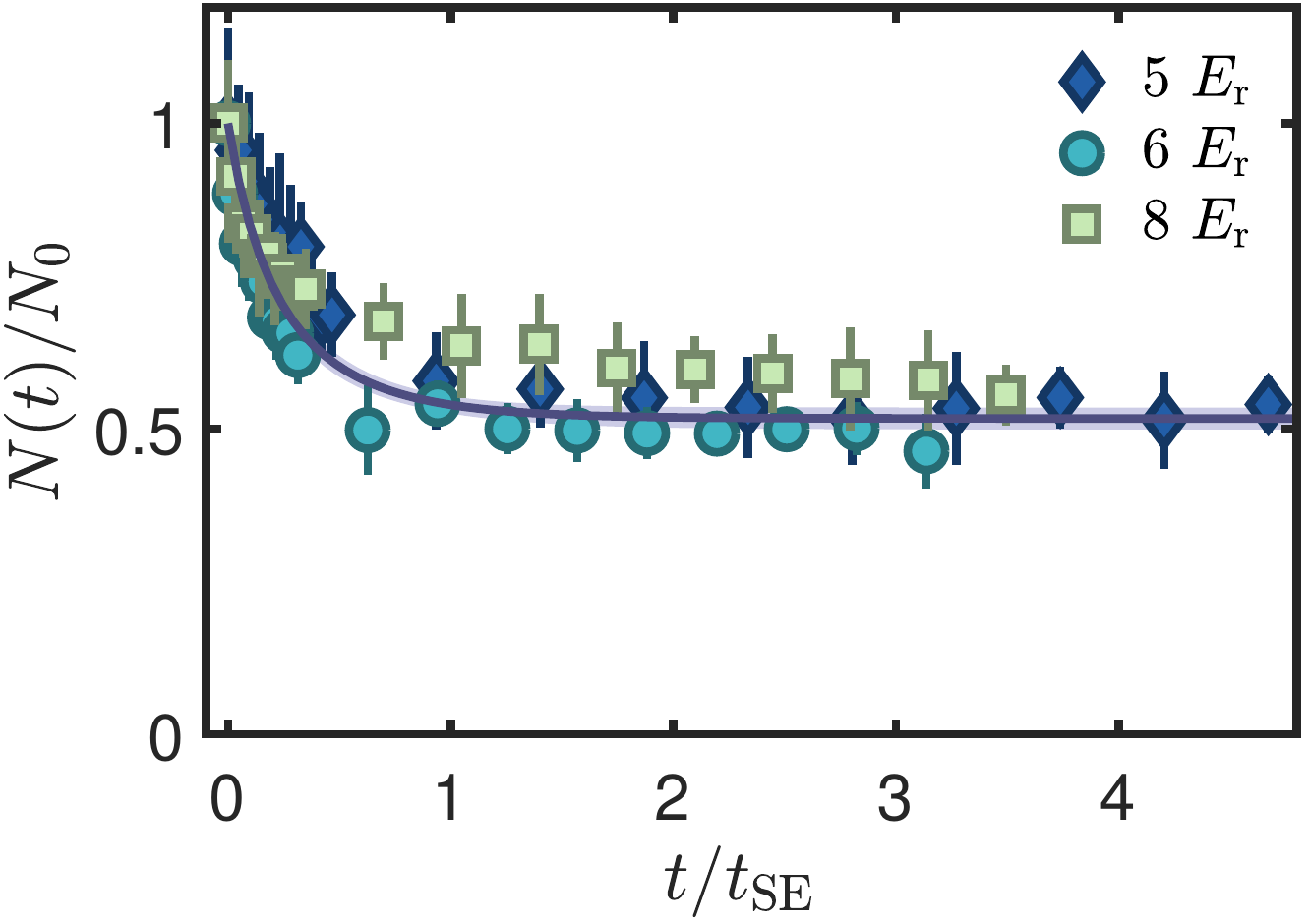}
  }
  \hfill
  \subfloat[\label{fig:norm_losses_568Er_6S}]
  {
    \includegraphics[width=0.45\textwidth]{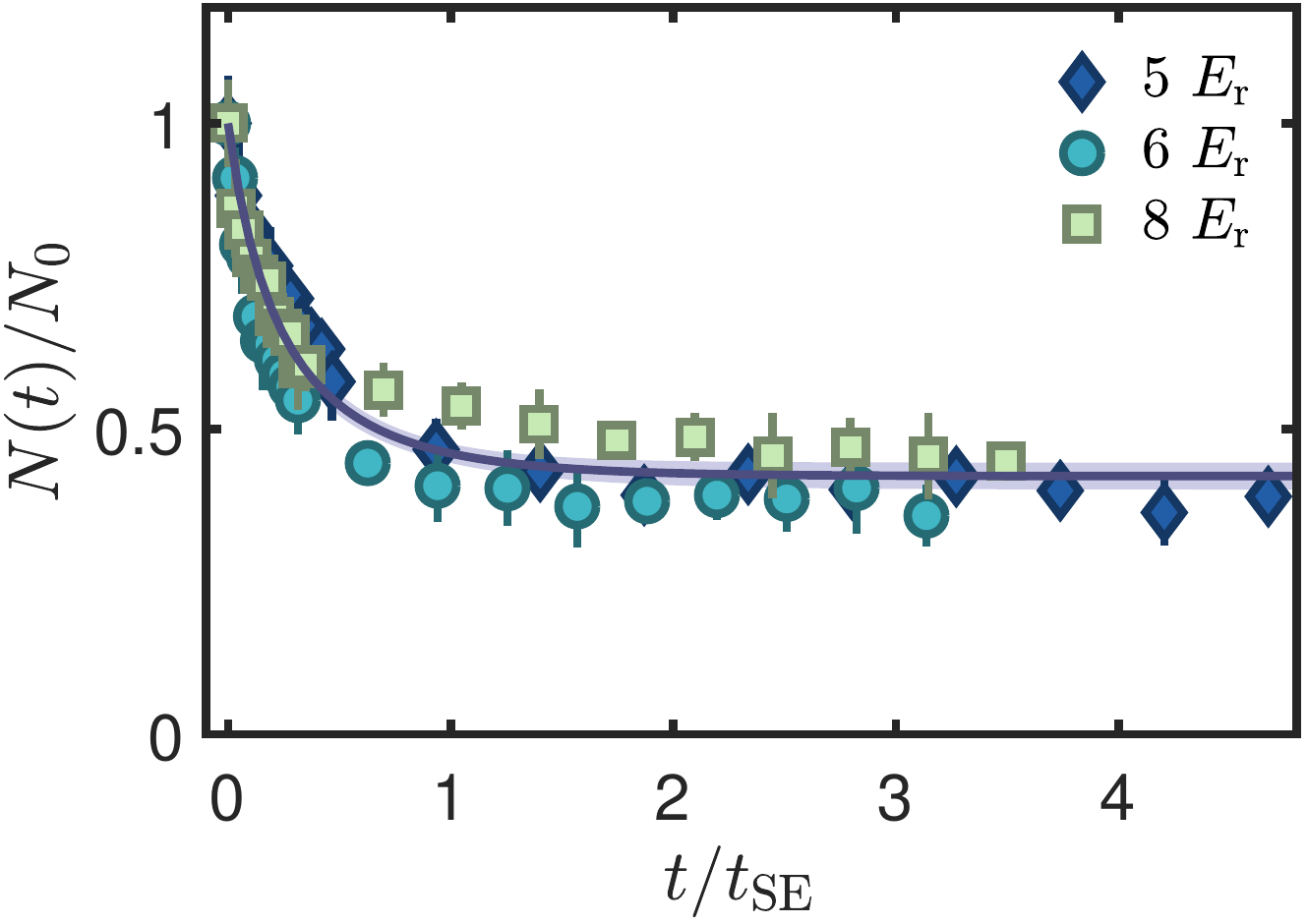}
  }
  \caption{Relative atom number $N(t)/N_0$ as a function of dimensionless time $t/t_\mathrm{SE}$ for (a) two- and (b) six-spin mixtures and for 5, 6 and $8~E_\mathrm{r}$ final lattice depth. The solid line shows the fit with decreasing $\kappa$ (see text) and the shaded area shows the 95\% confidence interval.}
  \label{fig:norm_losses_568Er_all}
\end{figure}

We repeat the loss measurements for several final lattice depths. 
As shown in \fref{fig:norm_losses_568Er_2S}, we observe essentially the same dynamics for all lattice depths after normalizing the data to a superexchange time $t_\mathrm{SE}=\hbar U /J^2$. 
After about one $t_\mathrm{SE}$, the two-body losses almost completely cease, indicating universal dynamics leading to $g^{(2)}(t \rightarrow \infty) \rightarrow 0$. 
Remarkably, we see no significant difference in the remaining fraction of atoms for different lattice depths.
We find good qualitative agreement with our numerical simulations shown in \fref{fig:Theory_atom_number_and_g2}, however the timescales of the transient dynamics do not compare perfectly, which requires further research beyond the scope this work. 


As mentioned earlier, there are several configurations of atoms in a lattice characterized by a vanishing $g^{(2)}$.
However, in \cite{Foss-Feig2012} it was shown that for a purely inelastically interacting 1D Fermi-Hubbard system ($U=0$) the final state of the time evolution, which has a finite remaining fraction of atoms, is uniquely determined and is in fact a highly entangled Dicke state, depicted in \fref{fig:bloch_sphere}.
In \cite{Foss-Feig2012} an expression for the final atom number was derived that only depends on the initial atom number, provided the initial state is an uncorrelated spin state.
These Dicke states, characterized by a completely antisymmetric spatial wave function, are dark with respect to the two-body inelastic collisions, which explains the absence of further particle loss in the final state.
Our numerical simulations, which are similar to \cite{Foss-Feig2012} but account for a finite $U$, reveal that the final state is independent of $U$.
\Fref{fig:Theory_Comp_vary_U} shows the simulation of the atom number dynamics for varying ratio $\hbar \Gamma_0/U$.
While the timescale for the transient dynamics changes, the final atom number is unaffected by the on-site interaction, and yields the same result as in \cite{Foss-Feig2012}.

To estimate the remaining fraction of atoms $f_\mathrm{th}$ for our system we first calculate the number of atoms per 1D tube and use this as a weighting to obtain the remaining fraction from the result of \cite{Foss-Feig2012}.
The measured remaining fraction of atoms for two-spin components $\bar{f}_2 = 0.51 \pm 0.03$, which we determine from the average of the last five data points, is however larger than the expected $f_\mathrm{th} \approx 0.27$ for an uncorrelated initial state. 

We mainly attribute the larger remaining fraction to two effects.
Firstly, residual ground-state atoms cut the 1D tubes into systems with effectively smaller atom number due to the sufficiently large on-site interaction $U_\mathrm{eg}$ between metastable-state atoms and ground-state atoms.
In our system we observe approximately 7\% ground-state atoms as a result of imperfect preparation. 
The expected remaining fraction is $f_\mathrm{th,g} \approx 0.35$, assuming the ground-state atoms are distributed randomly over the system.

Secondly, we attribute the observed larger remaining atom fraction to a higher Dicke state weight in the initial state at $t=0$, which can be understood as follows. 
Due to finite temperature and quantum fluctuations, the initial ground-state atoms Mott-insulator prior to preparation of the metastable-state contains a certain amount of doubly occupied sites.
Atoms on these doubly occupied sites are immediately lost by inelastic two-body collisions during metastable-state preparation via rapid adiabatic passage. 
Since only the symmetric part of the spatial wave function is affected by these losses, finite correlations, namely $g^{(2)} < 1$, already emerge before the start of the actual measurement at $t=0$.
We therefore expect a higher Dicke state weight in the initial state compared to a completely uncorrelated state. 

This higher initial Dicke state weight leads to a larger fraction of remaining atoms, which is also supported by our numerical simulations, where we have studied initial states with different correlations, as shown in \fref{fig:Theory_atom_number_and_g2}.
Interestingly, the quantum antiferromagnetic state, which we initialize in our simulations by taking the ground-state of the Fermi-Hubbard model, has a fully symmetric spatial wave function and exhibits a complete loss of atoms, as shown in \fref{fig:Theory_atom_number_and_g2}.
This potentially offers new possibilities for the experimental characterization of antiferromagnetic correlations at finite temperatures. 

We have further investigated how the suppression of losses depends on the number of spin states, namely in addition to the measurements in a two-spin system we have experimentally studied Fermi gases involving six different spin states. 
Our results are summarized in \fref{fig:norm_losses_568Er_6S}. 
The observed dynamics appears to be very similar to the $\mathrm{SU}(2)$ case. 
After a fast initial loss over a time of roughly $t_\mathrm{SE}$ the atom number stays constant and is again independent of the lattice depth. 
We find differences in the remaining fractions of atoms $\bar{f}_6~=~0.41~\pm~0.02$ and $\bar{f}_2~=~0.51~\pm~0.03$, which leads to $\bar{f}_6/\bar{f}_2~=~0.80~\pm~0.06$.
From fits of \eref{eq:solution_exp_dec_kappa} to the data sets shown in \fref{fig:norm_losses_568Er_all}, we extract the initial correlations $g^{(2)}_{0,6}/g^{(2)}_{0,2}~=~0.70~\pm~0.06$ and the respective timescales $\tau_6/\tau_2~=~1.25~\pm~0.14$ for the two different cases, and find indications for a non-trivial dependence of the build-up of correlations on the number of spin states. 
We attribute the difference in $g^{(2)}_0$ to our preparation via rapid adiabatic passage which works cleaner for the two-spin mixtures and thus imprints more initial correlations for six-spin mixtures.
While stronger initial correlations ($g^{(2)}<1$) would lead to a larger remaining fraction, we find a smaller remaining fraction for the six-spin mixtures.
We understand this based on the idea that a system with more spin components is less affected by the Pauli exclusion principle and thus exhibits stronger losses. This is also reflected in the right hand side of \eref{eq:solution_exp_dec_kappa} where the remaining fraction directly depends on $\mathcal{P}_\mathcal{N}$.

\section{Conclusions}
\label{sec:conclusions}

In conclusion we investigated dynamics in the one-dimensional dissipative Fermi-Hubbard model using two- and six-spin component $^{173}$Yb quantum gases.
We observe universal dynamics governed by an initial fast atom number decay driven by inelastic two-body collisions on the timescale of the superexchange.
Subsequently, we witness the formation of a strongly correlated many-body state that is stable with respect to further collisional loss on experimentally relevant time scales and is robust against experimental subtleties like initial particle number fluctuations or changes in the lattice depth.
Supported by existing predictions \cite{Foss-Feig2012} and our numerical simulations of the full master equation we find evidence that these states are likely to be highly entangled Dicke states.
These Dicke states could prove to be useful for future metrological applications.
By extending our measurements to a six-spin fermionic system, where we observe qualitatively similar behaviour, we underline that the formation of totally symmetric spin states by inelastic two-body losses seems to be a generic effect for $\mathrm{SU}(\mathcal{N})$ symmetric high-spin systems.
Our comparison of two- and six-spin state measurements reveals non-trivial differences in the time required to establish the correlated final state, as well as for the final atom number, which requires further theoretical studies.
Future experimental efforts could include tomographic \cite{Lucke2014,Apellaniz2015} or spectrographic \cite{Foss-Feig2012} measurements to unambiguously prove the generation of Dicke states. 
In addition, photo-association could be employed to tune $\hbar \Gamma_0 / U$ \cite{Tomita2017} and study loss dynamics across the phase diagram from the Mott insulator to the quantum Zeno insulator regime.

\ack
This work has been supported financially by the Deutsche Forschungsgemeinschaft within the SFB 925.
LF and LM acknowledge support from the excellence cluster `The Hamburg Centre for Ultrafast Imaging - Structure, Dynamics and Control of Matter at the Atomic Scale' of the Deutsche Forschungsgemeinschaft.
TP acknowledges funding from Marie Curie ITN Project QTea (MRTN-CT-2012-317485).

\appendix
\section{Master Equation}
\label{sec_app:master_equation}
We describe the dynamics of the system with a numerical implementation of the master equation \cite{Lindblad1976,Gorini}
\begin{equation}
  \hbar \frac{\mathrm{d}}{\mathrm{d}t}\rho (t)=-\mathrm{i}[H,\rho(t)] -\frac{1}{2}\sum_{i=1}^N\left(L_i^\dagger L_i^{} \rho(t) +\rho(t)L_i^\dagger L_i^{}  -2L_i^{} \rho(t)L_i^\dagger\right),
\end{equation}
which governs the time evolution of the density matrix $\rho(t)$. 
The dynamics has unitary and dissipative contributions, given by the first and second term, respectively. 
The Hamiltonian $H$ is the Fermi-Hubbard model
\begin{equation}
  H = U\sum_{i=1}^{N}\nu{i}\nd{i}-J\sum_{i=1}^{N-1}(\cud{i}\cu{i+1}^{} +\cdd{i}\cd{i+1}^{} +h.c.),
\end{equation}
where $\hat{c}_{i\sigma}^\dagger$ is the fermionic creation operator for site $i$ and spin $\sigma$. $\hat{n}_{i\sigma} \equiv \hat{c}_{i\sigma}^\dagger \hat{c}_{i\sigma}^{}$ denotes the number operator for a specific spin state. 
The dissipative contributions are given by the Lindblad operators $L_i = \sqrt{\hbar \Gamma_0}\cu{i}\cd{i}$.

We initialize the system in an incoherent mixture of states without double occupancy at half filling with uniform probability distribution for these states in the mixture. 
The most convenient basis for this initialization is the Fock basis.
However, to include initial correlations we need a basis that includes the six-site Dicke state given by $\ket{\Psi_D}=\sum_i P_i(\cud{1}\cud{2}\cud{3}\cdd{4}\cdd{5}\cdd{6})\ket{0}$, where $P_i$ denotes all possible permutations. 
Therefore, we use a Gram-Schmidt algorithm to create a complete basis set consisting of the dark Dicke state $\ket{\Psi_D}$, and 19 bright states $\ket{\Psi_{L,i}}$ which are orthogonal to $\ket{\Psi_D}$.
The initial density matrix for the simulations is then $\rho_I=w_D\rho_D+\sum_i w_{L}\rho_{L,i}$, where $\rho_\alpha\equiv\ket{\Psi_\alpha}\bra{\Psi_\alpha}$ for a state $\ket{\Psi_\alpha}$.
We adjust the weights $w_\alpha$ to study initial correlations of the system while still satisfying the normalization relation $w_D+19w_L=1$.

The simulation of dynamics of the quantum antiferromagnetic initial state is initialized with the ground-state of the Fermi-Hubbard model.

\section{Generalized Two-Body Loss Rate Equation}
\label{sec_app:general_2body_loss_eq}

We generalize the differential equation describing our observable $N(t)$ to $\mathcal{N}$ spin components. For a single spin component $m$, the loss rate equation is given by
\begin{equation}
 \frac{\rmd N_m}{\rmd t} = - \frac{\kappa}{N_0} N_m \sum_{m' \neq m} N_{m'},
\end{equation}
where $m \in \{1,\dots,\mathcal{N}\}$. Assuming equal distribution of spins $N_m = N_{m'}$ this gives
\begin{equation}
 \frac{\rmd N_m}{\rmd t} = - (\mathcal{N}-1) \frac{\kappa}{N_0} N_m^2.
\end{equation}
The total atom number $N = \sum_{m'}N_{m'} = \mathcal{N} N_m$ then yields
\begin{eqnarray}
 \frac{\rmd N}{\rmd t} & = - \frac{\mathcal{N}-1}{\mathcal{N}} \frac{\kappa}{N_0} N^2\\
 \frac{\rmd N}{\rmd t} & = - \mathcal{P}_\mathcal{N} \frac{\kappa}{N_0} N^2,
\end{eqnarray}
where we have defined $\mathcal{P}_\mathcal{N} \equiv (\mathcal{N}-1)/\mathcal{N}$.

\section{Nearest Neighbour Correlation Function for a Dicke State}
\label{sec_app:g2_Dicke_State}

The Dicke state predicted in \cite{Foss-Feig2012} is given by
\begin{equation}
 \Psi = \mathcal{A} \Phi(r_1, \dots, r_N) \sum_{\vec{\sigma}} \left| \vec{\sigma} \right>,
\end{equation}
where $\mathcal{A}$ is some normalization, $\Phi(r_1, \dots, r_N)$ is the fully antisymmetric spatial wave function, and $\sum_{\vec{\sigma}} \left| \vec{\sigma} \right> = \sum_{\sigma_1}\cdots\sum_{\sigma_N} \left| \sigma_1 \right> \otimes \cdots \otimes \left| \sigma_N \right>$ is the fully symmetric spin wave function. For convenience, we will write $\Phi(r_1, \dots, r_N)$ as $\left| \Phi \right>$.

We consider the numerator of \eref{eq:g2} for any pair of nearest neighbours
\begin{eqnarray}
& |\mathcal{A}|^2 \left(  \left< \Phi \right|  \sum_{\vec{\sigma}} \left< \vec{\sigma} \right| \hat{n}_i \hat{n}_j \left| \Phi \right>  \sum_{\vec{\sigma}'} \left| \vec{\sigma}' \right> -\frac{4}{\hbar^2} \left< \Phi \right|  \sum_{\vec{\sigma}} \left< \vec{\sigma} \right| \hat{\bi{S}}_i \cdot \hat{\bi{S}}_j \left| \Phi \right>  \sum_{\vec{\sigma}'} \left| \vec{\sigma}' \right>   \right) \\
= & |\mathcal{A}|^2 \left(    \sum_{\vec{\sigma}} \sum_{\vec{\sigma}'}\left< \vec{\sigma} | \vec{\sigma}' \right>  \left< \Phi \right|\hat{n}_i \hat{n}_j \left| \Phi \right>    -\frac{4}{\hbar^2} \underbrace{\left< \Phi | \Phi \right>}_{=1}  \sum_{\vec{\sigma}}\sum_{\vec{\sigma}'} \left< \vec{\sigma} \right| \hat{\bi{S}}_i \cdot \hat{\bi{S}}_j    \left| \vec{\sigma}' \right>   \right)\\
= & |\mathcal{A}|^2 B \left(   \left< \sigma_{i,j} | \sigma_{i,j}' \right>   -\frac{4}{\hbar^2}    \left< \sigma_{i,j} \right| \hat{\bi{S}}_i \cdot \hat{\bi{S}}_j    \left| \sigma_{i,j}' \right>   \right),
\end{eqnarray}
where we have introduced $B \equiv \sum_{\vec{\sigma} \neq \sigma_i, \sigma_j} \sum_{\vec{\sigma}' \neq \sigma_i', \sigma_j'} \left< \vec{\sigma} \neq \sigma_i, \sigma_j | \vec{\sigma}' \neq \sigma_i', \sigma_j'  \right>$ as the sum over all spin states except the nearest neighbours $i$ and $j$.
From (C.3) to (C.4) we have used that $\left< \Phi \right|\hat{n}_i \hat{n}_j \left| \Phi \right>$ is either 1 if there is an atom on both nearest neighbouring sites, or 0 if one or two of the nearest neighbouring sites is empty. 
In the latter case, the contribution to $g^{(2)}$ is zero, since then also the second term will yield zero. 
Further we have introduced $\left| \sigma_{i,j} \right> \equiv \sum_{\sigma_i} \sum_{\sigma_j} \left| \sigma_i \right> \otimes \left| \sigma_j \right>$, which can be written as $\left| \uparrow \uparrow \right> + \left| \uparrow \downarrow \right> + \left| \downarrow \uparrow \right> + \left| \downarrow \downarrow \right>$.
  
Equation (C.4) further yields
\begin{eqnarray}
& \left< \sigma_{i,j} | \sigma_{i,j}' \right>   -\frac{4}{\hbar^2}    \left< \sigma_{i,j} \right| \hat{\bi{S}}_i \cdot \hat{\bi{S}}_j    \left| \sigma_{i,j}' \right>  \\
= & 4   -\frac{4}{\hbar^2}  \left(  \left< \sigma_{i,j} \right| \frac{1}{2}\left( \hat{S}_{+,i}\hat{S}_{-,j} + \hat{S}_{-,i}\hat{S}_{+,j} \right)\left| \sigma_{i,j}' \right> + \underbrace{\left< \sigma_{i,j} \right|\hat{S}_{z,i} \hat{S}_{z,j}   \left| \sigma_{i,j}' \right>}_{=0}  \right) \\
= & 4   -\frac{2}{\hbar^2}    \left< \sigma_{i,j} \right| \left( \hat{S}_{+,i}\hat{S}_{-,j} + \hat{S}_{-,i}\hat{S}_{+,j} \right)\left| \sigma_{i,j}' \right>  \\
= & 4   -\frac{2}{\hbar^2} \left(   \underbrace{\left< \uparrow \downarrow \right|  \hat{S}_{+,i}\hat{S}_{-,j}\left|  \downarrow \uparrow \right>}_{=\hbar^2}  + \underbrace{\left< \downarrow \uparrow \right| \hat{S}_{-,i}\hat{S}_{+,j} \left| \uparrow \downarrow \right>}_{=\hbar^2} \right)\\
= & 0.
\end{eqnarray}
We therefore note that for each pair of nearest neighbours, the contribution to $g^{(2)}$ is zero, so summing over the whole system also yields $g^{(2)} = 0$ for the Dicke state.

\section*{References}
\bibliography{dissipation_bib}

\providecommand{\newblock}{}
\begin{thebibliography}{10}
\expandafter\ifx\csname url\endcsname\relax
  \def\url#1{{\tt #1}}\fi
\expandafter\ifx\csname urlprefix\endcsname\relax\def\urlprefix{URL }\fi
\providecommand{\eprint}[2][]{\url{#2}}

\bibitem{Diehl2008}
Diehl S, Micheli A, Kantian A, Kraus B, B{\"{u}}chler H~P and Zoller P 2008
  {\em Nature Physics\/} {\bf 4} 878--883 ISSN 17452473 (\textit{Preprint}
  \eprint{0803.1482})

\bibitem{Verstraete2009}
Verstraete F, Wolf M~M and {Ignacio Cirac} J 2009 {\em Nature Physics\/} {\bf
  5} 633--636 ISSN 17452473 (\textit{Preprint} \eprint{0803.1447})

\bibitem{Diehl2011}
Diehl S, Rico E, Baranov M~A and Zoller P 2011 {\em Nature Physics\/} {\bf 7}
  971--977 ISSN 17452481 (\textit{Preprint} \eprint{1105.5947})

\bibitem{Foss-Feig2012}
Foss-Feig M, Daley A~J, Thompson J~K and Rey A~M 2012 {\em Physical Review
  Letters\/} {\bf 109} 1--5 ISSN 00319007 (\textit{Preprint}
  \eprint{arXiv:1207.4741v1})

\bibitem{Griessner2006}
Griessner A, Daley A~J, Clark S~R, Jaksch D and Zoller P 2006 {\em Physical
  Review Letters\/} {\bf 97} 1--4 ISSN 00319007 (\textit{Preprint}
  \eprint{0607254})

\bibitem{Baur2010}
Baur S~K and Mueller E~J 2010 {\em Physical Review A - Atomic, Molecular, and
  Optical Physics\/} {\bf 82} ISSN 10502947 (\textit{Preprint}
  \eprint{1003.5235})

\bibitem{Gross2017}
Gross C and Bloch I 2017 {\em Science\/} {\bf 357} 995--1001 ISSN 0036-8075
  (\textit{Preprint}
  \eprint{http://science.sciencemag.org/content/357/6355/995.full.pdf})

\bibitem{Catani2009}
Catani J, Barontini G, Lamporesi G, Rabatti F, Thalhammer G, Minardi F,
  Stringari S and Inguscio M 2009 {\em Physical Review Letters\/} {\bf 103}
  2--5 ISSN 00319007 (\textit{Preprint} \eprint{0906.2264})

\bibitem{Chen2014}
Chen D, Meldgin C and Demarco B 2014 {\em Physical Review A - Atomic,
  Molecular, and Optical Physics\/} {\bf 90} 1--5 ISSN 10941622
  (\textit{Preprint} \eprint{1401.5096})

\bibitem{Barontini2013}
Barontini G, Labouvie R, Stubenrauch F, Vogler A, Guarrera V and Ott H 2013
  {\em Physical Review Letters\/} {\bf 110} 1--5 ISSN 00319007
  (\textit{Preprint} \eprint{arXiv:1212.4824v1})

\bibitem{Labouvie2015}
Labouvie R, Santra B, Heun S, Wimberger S and Ott H 2015 {\em Physical Review
  Letters\/} {\bf 115} 1--5 ISSN 10797114 (\textit{Preprint}
  \eprint{1411.5632v1})

\bibitem{Labouvie2016}
Labouvie R, Santra B, Heun S and Ott H 2016 {\em Physical Review Letters\/}
  {\bf 116} 1--5 ISSN 10797114

\bibitem{Luschen2017}
L{\"{u}}schen H~P, Bordia P, Hodgman S~S, Schreiber M, Sarkar S, Daley A~J,
  Fischer M~H, Altman E, Bloch I and Schneider U 2017 {\em Physical Review X\/}
  {\bf 7} 1--13 ISSN 21603308 (\textit{Preprint} \eprint{1610.01613})

\bibitem{Patil2015}
Patil Y~S, Chakram S and Vengalattore M 2015 {\em Physical Review Letters\/}
  {\bf 115} 1--5 ISSN 10797114 (\textit{Preprint} \eprint{1411.2678})

\bibitem{Syassen2008}
Syassen N, Bauer D~M, Lettner M, Volz T, Dietze D, Garc{\'{i}}a-Ripoll J~J,
  Cirac J~I, Rempe G and D{\"{u}}rr S 2008 {\em Science\/} {\bf 320} 1329--1331
  ISSN 00368075 (\textit{Preprint} \eprint{0806.4310})

\bibitem{Garcia-Ripoll2009}
Garc{\'{i}}a-Ripoll J~J, D{\"{u}}rr S, Syassen N, Bauer D~M, Lettner M, Rempe G
  and Cirac J~I 2009 {\em New Journal of Physics\/} {\bf 11} ISSN 13672630

\bibitem{Durr2009}
D{\"{u}}rr S, Garc{\'{i}}a-Ripoll J~J, Syassen N, Bauer D~M, Lettner M, Cirac
  J~I and Rempe G 2009 {\em Physical Review A - Atomic, Molecular, and Optical
  Physics\/} {\bf 79} 1--13 ISSN 10502947 (\textit{Preprint}
  \eprint{arXiv:0809.3696v2})

\bibitem{Yan2013}
Yan B, Moses S~A, Gadway B, Covey J~P, Hazzard K~R~A, Rey A~M, Jin D~S and Ye J
  2013 {\em Nature\/} {\bf 501} 521--525 ISSN 1476-4687 (\textit{Preprint}
  \eprint{1305.5598})

\bibitem{Zhu2014}
Zhu B, Gadway B, Foss-Feig M, Schachenmayer J, Wall M~L, Hazzard K~R~A, Yan B,
  Moses S~A, Covey J~P, Jin D~S, Ye J, Holland M and Rey A~M 2014 {\em Physical
  Review Letters\/} {\bf 112} 1--5 ISSN 00319007 (\textit{Preprint}
  \eprint{1310.2221})

\bibitem{Tomita2017}
Tomita T, Nakajima S, Danshita I, Takasu Y and Takahashi Y 2017   1--9
  (\textit{Preprint} \eprint{1705.09942})

\bibitem{Mark2012}
Mark M~J, Haller E, Lauber K, Danzl J~G, Janisch A, B{\"{u}}chler H~P, Daley
  A~J and N{\"{a}}gerl H~C 2012 {\em Physical Review Letters\/} {\bf 108} 1--5
  ISSN 00319007 (\textit{Preprint} \eprint{arXiv:1108.3013v1})

\bibitem{Radiation1954}
Dicke R 1954 {\em Physical Review\/} {\bf 93} 99--110 ISSN 0031-899X
  (\textit{Preprint} \eprint{1407.7336})

\bibitem{Lucke2014}
L{\"{u}}cke B, Peise J, Vitagliano G, Arlt J, Santos L, T{\'{o}}th G and Klempt
  C 2014 {\em Physical Review Letters\/} {\bf 112} 1--5 ISSN 10797114
  (\textit{Preprint} \eprint{1403.4542})

\bibitem{Apellaniz2015}
Apellaniz I, L{\"{u}}cke B, Peise J, Klempt C and T{\'{o}}th G 2015 {\em New
  Journal of Physics\/} {\bf 17} ISSN 13672630 (\textit{Preprint}
  \eprint{1412.3426})

\bibitem{Scazza2014}
Scazza F, Hofrichter C, H{\"{o}}fer M, {De Groot} P~C, Bloch I and
  F{\"{o}}lling S 2014 {\em Nature Physics\/} {\bf 10} 779--784 ISSN 17452481
  (\textit{Preprint} \eprint{1403.4761})

\bibitem{Kitagawa2008}
Kitagawa M, Enomoto K, Kasa K, Takahashi Y, Ciury{\l}o R, Naidon P and Julienne
  P~S 2008 {\em Physical Review A - Atomic, Molecular, and Optical Physics\/}
  {\bf 77} ISSN 10502947 (\textit{Preprint} \eprint{0708.0752})

\bibitem{Taie2010}
Taie S, Takasu Y, Sugawa S, Yamazaki R, Tsujimoto T, Murakami R and Takahashi Y
  2010 {\em Physical Review Letters\/} {\bf 105} 1--4 ISSN 00319007

\bibitem{Krauser2012}
Krauser J~S, Heinze J, Fl{\"{a}}schner N, G{\"{o}}tze S, J{\"{u}}rgensen O,
  L{\"{u}}hmann D~S, Becker C and Sengstock K 2012 {\em Nature Physics\/} {\bf
  8} 813--818 ISSN 17452473 (\textit{Preprint} \eprint{1203.0948})

\bibitem{Stamper-Kurn2013}
Stamper-Kurn D~M and Ueda M 2013 {\em Reviews of Modern Physics\/} {\bf 85}
  1191--1244 ISSN 00346861 (\textit{Preprint} \eprint{1205.1888})

\bibitem{Krauser2014}
Krauser J~S, Ebling U, Fl{\"{a}}schner N, Heinze J, Sengstock K, Lewenstein M,
  Eckardt A and Becker C 2014 {\em Science\/} {\bf 343} 157--160 ISSN 10959203
  (\textit{Preprint} \eprint{1307.8392})

\bibitem{Pagano2014}
Pagano G, Mancini M, Cappellini G, Lombardi P, Sch{\"{a}}fer F, Hu H, Liu X~J,
  Catani J, Sias C, Inguscio M and Fallani L 2014 {\em Nature Physics\/} {\bf
  10} 198--201 ISSN 17452473 (\textit{Preprint} \eprint{1408.0928})

\bibitem{Hofrichter2016}
Hofrichter C, Riegger L, Scazza F, H{\"{o}}fer M, Fernandes D~R, Bloch I and
  F{\"{o}}lling S 2016 {\em Physical Review X\/} {\bf 6} ISSN 21603308
  (\textit{Preprint} \eprint{1511.07287})

\bibitem{Dorscher2013}
D{\"{o}}rscher S, Thobe A, Hundt B, Kochanke A, {Le Targat} R, Windpassinger P,
  Becker C and Sengstock K 2013 {\em Review of Scientific Instruments\/} {\bf
  84} ISSN 00346748 (\textit{Preprint} \eprint{1303.1105})

\bibitem{Lindblad1976}
Lindblad G 1976 {\em Communications in Mathematical Physics\/} {\bf 48}
  119--130 ISSN 00103616 (\textit{Preprint} \eprint{arXiv:1402.7287v1})

\bibitem{Gorini}
Gorini V, Kossakowski A and Sudarshan E~C~G 1976 {\em Journal of Mathematical
  Physics\/} {\bf 17} 821--825 (\textit{Preprint}
  \eprint{https://aip.scitation.org/doi/pdf/10.1063/1.522979})

\end{thebibliography}

\end{document}